\documentclass[pre,aps,twocolumn,showpacs]{revtex4}
\usepackage{graphicx}
\begin{document}

\title{Force correlations in the $q$--model for general $q$--distributions} 

\author{Jacco H. Snoeijer and J. M. J. van Leeuwen}
\affiliation{Instituut--Lorentz, Leiden University, P.O. Box  9506, 2300 RA Leiden, 
The Netherlands}

\date{\today} 
\begin{abstract}
We study force correlations in the $q$--model for granular media at infinite depth, 
for general $q$--distributions. We show that there are no 2--point force correlations 
as long as $q$--values at different sites are uncorrelated. However, higher order 
correlations can persist, and if they do, they only decay with a power of the distance. 
Furthermore, we find the entire set of $q$--distributions for which the force distribution factorizes. 
It includes distributions ranging from infinitely sharp to almost critical. Finally, we show 
that 2--point force correlations do appear whenever there are correlations between $q$--values 
at different sites in a layer; various cases are evaluated explicitly. 
\end{abstract}

\pacs{02.50.Ey, 45.70.Cc, 81.05.Rm}

\maketitle
\section{Introduction} \label{intro}
One of the main challenges of granular media is to characterize the network of 
microscopic forces in a static bead pack. In order to describe the 
corresponding force fluctuations, Liu et al. \cite{liu} introduced the $q$--model. In this model, 
the beads are placed on a regular lattice and the (scalar) forces are stochastically transmitted, 
by random fractions denoted by the symbol $q$. Even in its simplest version, where one assumes a 
uniform $q$--distribution, it already reproduces the main feature of the experimental observations: 
the probability for large forces decays exponentially \cite{liu,exp1,exp2}. Although for this uniform 
$q$--distribution the forces become totally uncorrelated, in general, correlations do persist 
\cite{cop}. In the present study, we investigate for which $q$--distributions this is the case 
and we reveal the surprising nature of these correlations. In order to perform an analytical 
study, we restrict ourselves to the scalar $q$--model and allow only correlations between 
$q$--values in a layer. More sophisticated lattice models, which include the vector nature 
of the force and allow correlations between layers are not considered here \cite{vec}.

Although the $q$--model is particularly simple, its behavior turns out to be very rich. First of all, 
there is a so-called critical $q$--distribution, which produces a force distribution that decays 
algebraically instead of exponentially \cite{cop,bouch}. It therefore forms a critical point 
in the space of $q$--distributions, and its properties were recently investigated in great 
detail \cite{raj,lew}. A second intriguing issue concerns the top--down {\em dynamics} of force 
correlations (the downward direction can be interpreted as time) \cite{raj,lew,jac}. Even if 
both in the initial state (top layer) and in the asymptotic state (infinite depth) all forces are 
uncorrelated, there will be correlations at all intermediate levels. Correlations become longer 
in range while their amplitudes diminish in a diffusion process, and as a result, the asymptotic 
force distribution is only approached algebraically \cite{jac}. This process is closely related 
to the subject of this study, namely the presence of force correlations at infinite depth. 

\begin{figure}[ht]
\includegraphics[width=4.2cm]{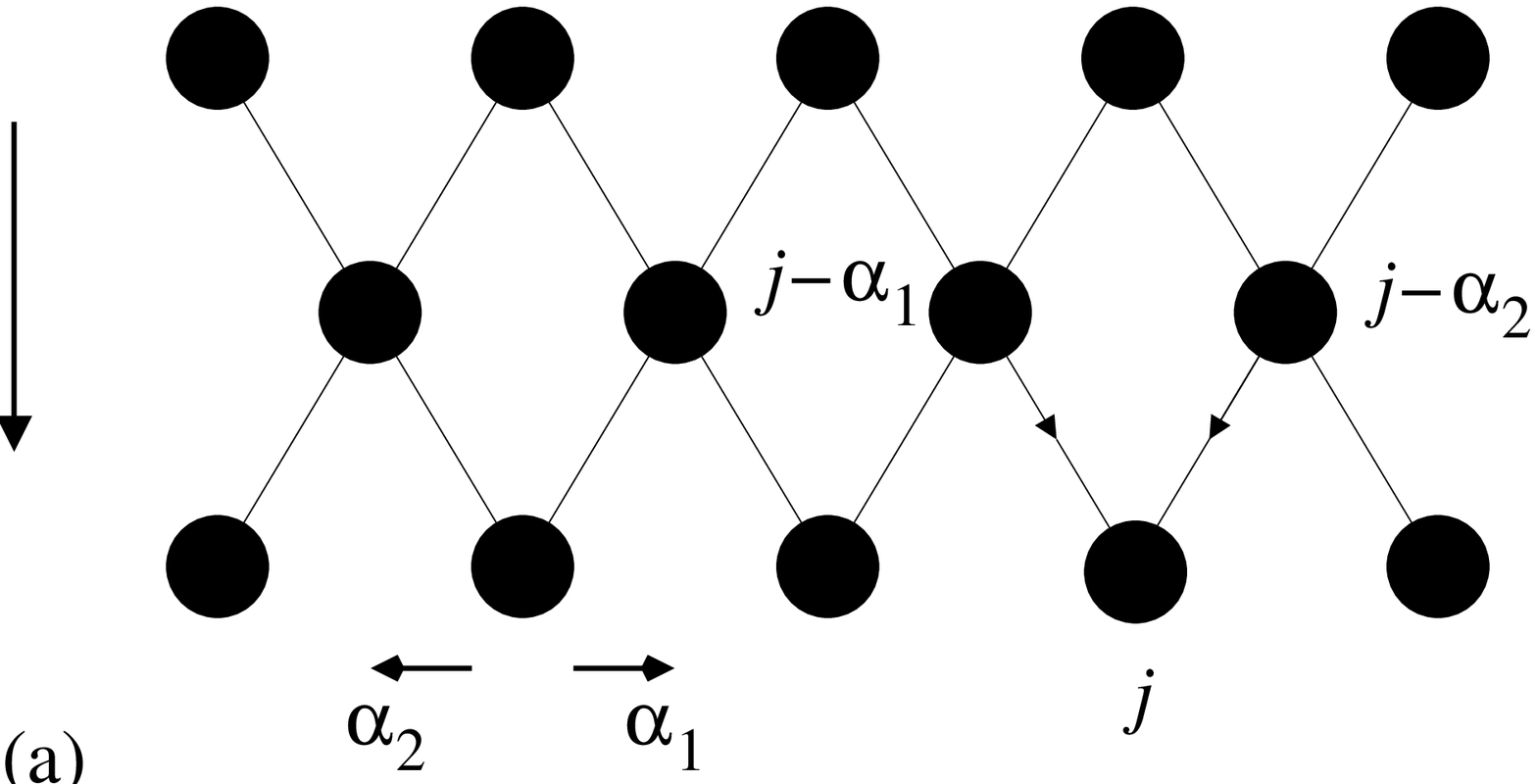}
\includegraphics[width=4.2cm]{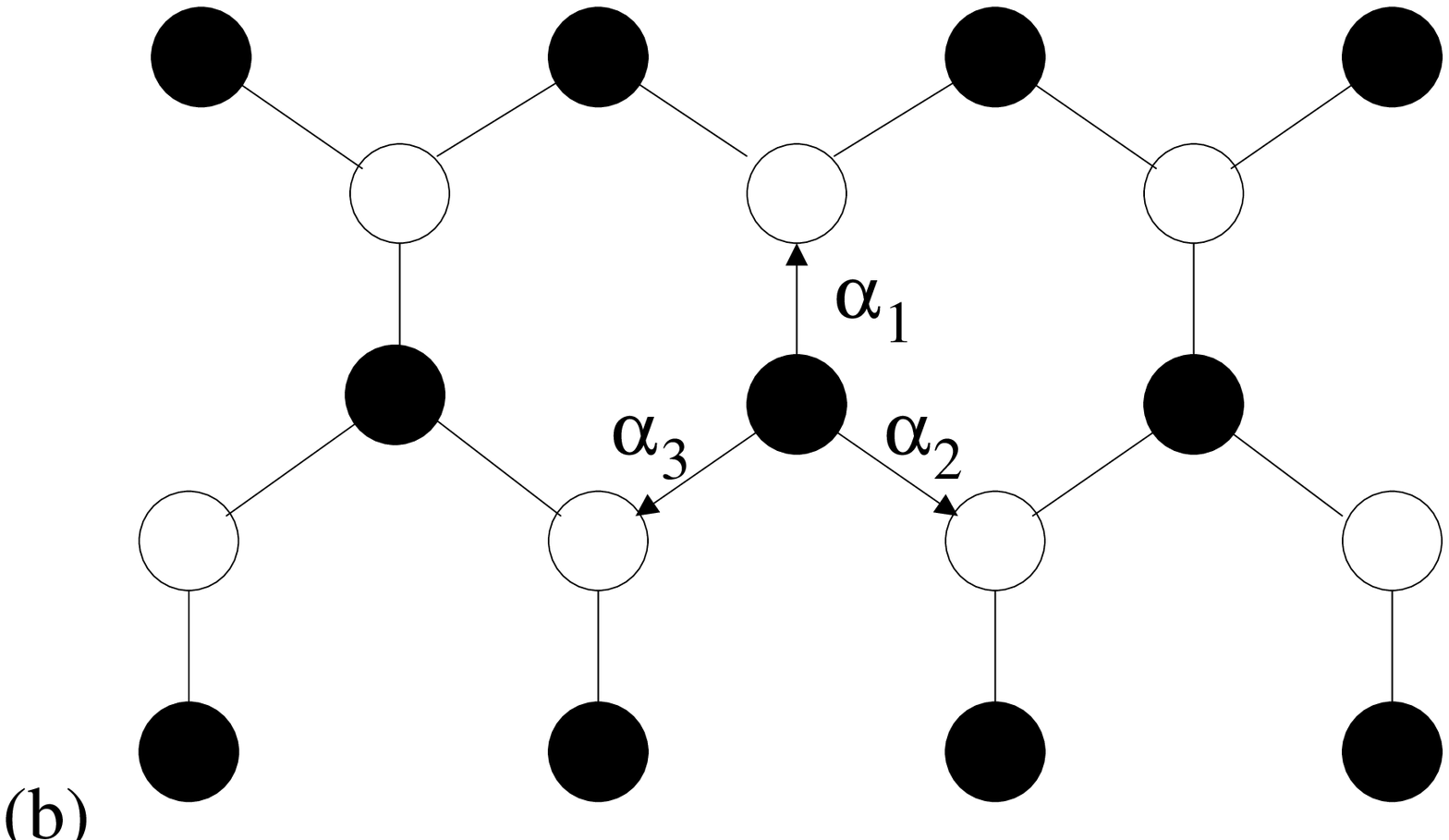}
\caption{The displacement vectors $\alpha$ in the $q$--model for (a) the triangular packing 
(side view) and (b) the fcc packing (top view)}
\label{qmd}
\end{figure}
Let us recapitulate the definition of the $q$--model. The beads are assumed to be 
positioned on a regular lattice. Let $f_i$ be the force in the downward 
direction on the $i$th bead in a layer. This bead makes contact with a number of $z$ beads 
in the layer below, which we indicate by the indices $i+\alpha$. The $\alpha$'s are displacement 
vectors in the lower layer as shown in Fig. \ref{qmd}. Bead $i$ transmits a fraction 
$q_{i,\alpha}$ of the force $f_i$ to the bead $i+\alpha$ underneath it. These fractions are 
taken stochastically from a distribution satisfying the constraint
\begin{equation} \label{a1}
\sum_\alpha q_{i,\alpha} = 1,
\end{equation}
which assures mechanical equilibrium in the vertical direction. So, we can write the force 
$f'_j$ on the $j$th bead in a layer as
\begin{equation} \label{a2}
f'_j = \sum_\alpha q_{j-\alpha,\alpha} \,f_{j-\alpha}.
\end{equation}
As the weights of the particles are unimportant at infinite depth, we have left out the 
so--called injection term. 
The distribution of forces at infinite depth depends on the $q$--distribution $H(\vec{q})$, where the 
symbol $\vec{q}$ is a shorthand for all the $q_{i,\alpha}$ at a given layer. This $H(\vec{q})$ 
can be any function that is constrained by Eq.(\ref{a1}).  
If we now assume that there are no correlations between the $q$--values at different sites, the 
$q$--distribution is of the form
\begin{equation}\label{a3}
H(\vec{q})=\prod_{i} \,\eta(\vec{q}_i)\, \delta \left(1-\sum_\alpha q_{i,\alpha} \right)\, , 
\quad \quad \vec{q}_i=\{q_{i,\alpha} \},
\end{equation}
where $\eta(\vec{q}_i)$ is symmetric in its arguments $q_{i,\alpha}$. Although we will refer to 
these $q$--distributions as ``uncorrelated'', note that there are always correlations between 
the $q_{i,\alpha}$ of the same site due to the $\delta$--constraint. 

In the first part of this 
study, we show that there is only a limited set of $\eta(\vec{q}_i)$ for which the stationary force 
distribution can be written as a product of single--site distributions, and therefore is totally 
uncorrelated. This set is an extension of the set that was already identified by Coppersmith 
et al. \cite{cop}. In their extensive study, they also provided numerical evidence that, in general, 
correlations can persist. We will show that correlations are still absent in the second order 
moments. However, higher order correlations do exist and surprisingly enough, these turn out 
to decay algebraically. 
The results for the triangular packing and the fcc packing are summarized in 
Table \ref{results}, section \ref{discussion}. 
In the last part of this work, we show that one induces 2--point force 
correlations by allowing correlations between $q$--values on  different sites in a layer. 
These correlations will generically vanish with a power law, except for the triangular 
packing, where the decay of force correlations follows the decay of the $q$--correlations.

The paper is organized as follows. In section \ref{criterion} we derive a criterion that a 
distribution $\eta(\vec{q}_i)$ has to obey in order to produce an uncorrelated stationary state. 
We then show in section \ref{factorization}, that this criterion is only obeyed for a limited 
set of $\eta(\vec{q_i})$. After that, we study the nature of the correlations, 
by writing the evolution of the force moments as master equations in section \ref{moments}, and by 
analysing the stationary solutions of these equations in section \ref{long--range}. Section \ref{correlq} 
deals with the effects of allowing correlations between the $\vec{q_i}$ of different sites in a layer, 
and the paper closes with a discussion.

\section{Criterion for factorization}\label{criterion}
Using the recursive nature of the force transmission, Eq.(\ref{a2}), one can write down the 
following recursive relation for the force distribution \cite{cop,jac}:
\begin{eqnarray} \label{b1}
P' (\vec{f}') &=& \int H(\vec{q}) d \vec{q} \,\int P (\vec{f}) d \vec{f} \nonumber \\ 
&& \times \prod_j \, \delta (f'_j - \sum_\alpha q_{j-\alpha,\alpha} \, f_{j-\alpha}),
\end{eqnarray}
where we have introduced a vector notation for the forces in one layer 
$\vec{f} = (f_1, \cdots , f_N)$, and for the integrations we use the abbreviations
\begin{equation} \label{sub2}
\int d \vec{f} = \prod_i \int^\infty_0 df_i \, , 
\end{equation}
\begin{equation} \label{sub2b}
\int d \vec{q} = \prod_i \, \int d \vec{q}_i = \prod_i \,\prod_\alpha \int^1_0 d q_{i,\alpha} .
\end{equation}
It is often convenient to work with the Laplace transform of Eq.(\ref{b1}). Defining the Laplace 
transform as
\begin{equation} \label{b2}
\tilde{P} (\vec{s}) = \int d \vec{f} \exp( -\vec{s} \cdot \vec{f}) \, P (\vec{f}),
\end{equation}
the recursion simplifies to \cite{cop,jac}
\begin{equation} \label{b3}
\tilde{P}' (\vec{s}) = \int H(\vec{q})d \vec{q} \, \tilde{P} (\vec{s} (\vec{q})),
\end{equation}
with
\begin{equation} \label{b4}
s_i (\vec{q}) = \sum_\alpha q_{i,\alpha} \,s_{i+\alpha}. 
\end{equation}
The two representations Eq.(\ref{b1}) and (\ref{b3}) are equivalent, and they will both be used, 
depending on the nature of the problem.

The force distribution at infinite depth $P^*(\vec{f})$ or $ \tilde{P}^* (\vec{s})$ can be obtained 
by finding the fixed point of the recursive relation. 
The main question of this section is to determine whether a given 
$H(\vec{q})$ leads to a $P^*(\vec{f})$ that is simply a product of single--site force 
distributions $p^*(f_i)$. In section \ref{correlq} we will show that this can only be the case 
for $q$--distributions of the type Eq.(\ref{a3}). So for this section, the question is: which 
$\eta(\vec{q_i})$ lead to uncorrelated asymptotic states?

To answer this question, let us assume that such a fixed point exists, i.e.
\begin{equation} \label{b5}
P^*(\vec{f})=\prod_i \, p^*(f_i),\quad\quad {\rm or} \quad\quad 
\tilde{P}^*(\vec{s})=\prod_i \,\tilde{p}^*(s_i). 
\end{equation}
Inserting this Ansatz into the Laplace representation of the recursion relation, Eq.(\ref{b3}), yields
\begin{eqnarray} \label{b6}
\tilde{P}^*(\vec{s})&=& \nonumber \\
\prod_i  \int &\!\!\! \eta(\vec{q}_i) \!\!\!& \delta 
\left( 1-\sum_\alpha q_{i,\alpha} \right) d \vec{q}_i \,\,\tilde{p}^*
\left( \sum_\alpha q_{i,\alpha} s_{i+\alpha} \right) \nonumber \\
&=&\prod_i \tilde{\psi}(s_{i+\alpha_1},\cdots,s_{i+\alpha_z}),
\end{eqnarray}
where the function $\tilde{\psi}(s_{i+\alpha_1},\cdots,s_{i+\alpha_z})$ is the outcome 
of the integral over the $\vec{q_i}$. The arguments represent the $z$ sites that are connected to 
site $i$ in the previous layer. Integrating out all forces except those at the $z$ sites connected to $i$ 
means putting all $s_j=0$ except the set $\{s_{i+\alpha}\}$:
\begin{eqnarray}\label{b6new}
\tilde{P}^*(s_{i+\alpha_1},\cdots,s_{i+\alpha_z}) &=& 
\tilde{\psi}(s_{i+\alpha_1},\cdots,s_{i+\alpha_z}) \nonumber \\
&& \!\! \times\prod_\alpha \tilde{\psi}(s_{i+\alpha},0,\cdots)^{z-1}.
\end{eqnarray}
This projection of the total force distribution can only factorize if 
$\tilde{\psi}(s_{i+\alpha_1},\cdots,s_{i+\alpha_z})$ is a product function as well, i.e.
\begin{equation} \label{b7}
\tilde{\psi}(s_{i+\alpha_1},\cdots,s_{i+\alpha_z})=\prod_\alpha \tilde{\psi}(s_{i+\alpha}).
\end{equation}
This leads to the following criterion for asymptotic factorization:
\begin{itemize}
\item Given a $q$--distribution 
$\eta(\vec{q})$, one can construct a factorized fixed point if, and only if, there is a function 
$\tilde{\psi}(s)$ that satisfies the following condition:
\begin{equation} \label{b8}
\int \eta(\vec{q}) \delta 
\left( 1-\sum_\alpha q_\alpha \right) d \vec{q}\left[ \tilde{\psi}
\left( \sum_\alpha q_\alpha s_\alpha \right) \right]^z \,=\, \prod_\alpha 
\tilde{\psi}(s_\alpha) .
\end{equation}
\item This function $\tilde{\psi}(s)$ is related to the single--site distribution as
\begin{equation}\label{b9}
\tilde{p}^*(s)=\left[ \tilde{\psi}(s) \right]^z .
\end{equation}
\end{itemize}
Here, we omitted the site index $i$, and furthermore, our formulation depends only on $z$ 
(the number of $q$--values per site) and not on the details of the lattice.

\section{Special class of $q$--distributions leading to factorization}\label{factorization}
It is a well--known fact that the so--called uniform distribution, in which $\eta(\vec{q}_i)$ is 
a constant, 
produces an uncorrelated asymptotic force distribution. In fact, Coppersmith et al. identified 
a countable set of $q$--distributions, of which the uniform distribution is a member, that 
have this property \cite{cop}. Although it might seem obvious that a uniform distribution leads to 
an uncorrelated asymptotic state, it is really not trivial. Due to the constraint of Eq.(\ref{a1}), 
there are correlations between the $q_{i,\alpha}$ on each site $i$, which induce force correlations 
that only disappear under the special conditions discussed in the previous section, Eq.(\ref{b8}).
In this section, we will show when these special conditions are obeyed.

There is a mathematical relation that is extremely important 
for the $q$--model \cite{zinn}:
\begin{eqnarray} \label{zinnjust}
\prod_\alpha {1 \over (1+ s_\alpha)^r} &=& {\Gamma (zr) \over [\,\Gamma (r)\,]^z }  
\int \,d\vec{q}  \,\,\delta \left(1-\sum_\alpha q_\alpha\right) \nonumber \\
&&\times \,\prod_\alpha (q_\alpha)^{r-1}  
\,{1 \over (1 + \sum_\alpha q_\alpha s_\alpha)^{zr}}\,. \nonumber \\
&&
\end{eqnarray}
It holds for any real $r > 0$. From this relation, it is immediately clear 
that for all $q$--distributions of the type
\begin{equation} \label{c2}
\eta(\vec{q})={\Gamma (zr) \over [\,\Gamma (r)\,]^z } \,\prod_\alpha (q_\alpha)^{r-1}\,,
\quad\quad r>0
\end{equation}
there is a $\tilde{\psi}(s)$ that obeys Eq.(\ref{b8}), namely
\begin{equation} \label{c3}
\tilde{\psi}(s)={1 \over (1+ s)^r}.
\end{equation}
The corresponding single--site force distributions are
\begin{equation} \label{c4}
\tilde{p}^*(s)={1 \over (1+ s/zr)^{zr}} \quad {\rm or} \quad
p^*(f)=\frac{(zr)^{zr}}{\Gamma (zr)} \, f^{zr-1}\, e^{-zrf}.
\end{equation}
We rescaled the Laplace variable $s$, in order to put $\langle f \rangle=1$. 
Coppersmith et al. already found these $q$--distributions for integer values of $r$, also 
based on Eq.(\ref{zinnjust}) \cite{cop}. However, it holds for any real $r>0$. 
This means that the set for which the stationary force distribution factorizes 
is substantially larger; it ranges from the infinitely sharp distribution ($r \rightarrow \infty$) 
to the critical distribution ($r \rightarrow 0$) \cite{foot3}. 
Note that one recovers the results for the uniform distribution by putting $r=1$.

Although there is a huge variety of $q$--distributions that lead to uncorrelated force 
distributions, in general one cannot find a $\tilde{\psi}(s)$ that obeys Eq.(\ref{b8}). 
We will prove this by making a Taylor expansion of $\tilde{\psi}(s)$ 
\begin{equation} \label{c5}
\tilde{\psi}(s)=\sum_{n=0}^\infty \psi_n \,s^n,
\end{equation}
and then try to solve for the coefficients $\psi_n$ by imposing Eq.(\ref{b8}). It turns out 
that the equations can only be solved under special conditions, which are precisely obeyed by  
the class of $q$--distributions given by Eq.(\ref{c2}). 

Let us first focus on the left hand side (LHS) of Eq.(\ref{b8}). 
The Taylor expansion will give rise to terms of the type 
$(q_1s_1)^{n_1}(q_2s_2)^{n_2} \cdots (q_zs_z)^{n_z}$,  
which have to be integrated over all $q_\alpha$. This leads to terms $s_1^{n_1}s_2^{n_2}...s_z^{n_z}$ 
with prefactors given by the {\em moments} of $\eta(\vec{q})$
\begin{equation} \label{c6}
\overline{q_1^{n_1}q_2^{n_2}...q_z^{n_z}} = \int \eta(\vec{q}) \delta 
\left( 1-\sum_\alpha q_\alpha \right) d \vec{q} \,\, q_1^{n_1}q_2^{n_2}...q_z^{n_z}.
\end{equation}
These moments are not independent, due to the constraint Eq.(\ref{a1}). 
In appendix \ref{qmoments}, we show that the moments
\begin{equation} \label{c7}
\eta_n=\int \eta(\vec{q}) \delta \left( 1-\sum_\alpha q_\alpha \right) d\vec{q} \,\, q_1^n
\end{equation}
are in fact sufficient to characterize all relevant moments of Eq.(\ref{c6}). 
Besides the moments, there are of course additional prefactors 
consisting of combinations of the $\psi_n$; these 
are the quantities we try to find, for a given $q$--distribution $\eta(\vec{q})$.

The right hand side (RHS) of Eq.(\ref{b8}) also produces terms 
$s_1^{n_1}s_2^{n_2}\cdots s_z^{n_z}$, 
with prefactors $\psi_{n_1}\psi_{n_2}\cdots \psi_{n_z}$. The remaining task is to equate the 
prefactors of the terms $s_1^{n_1}s_2^{n_2}\cdots s_z^{n_z}$ on both sides of the equation. 
This gives a set of equations, from which one can try to solve for the $\psi_n$. 

The zeroth order equation is trivially obeyed for any $\psi_0$, as can be seen by putting 
all $s_\alpha=0$. For convenience we fix $\psi_0=1$. The same happens at first order, 
since for each $\alpha$, the LHS contains $z$ terms 
$\psi_1 \overline{q_\alpha}s_\alpha = 1/z \,\psi_1 s_\alpha$, and the RHS is 
simply $\psi_1 s_\alpha$. The first non-trivial equation appears at second order. There are two 
equations, for $s_\alpha^2$ and for $s_{\alpha}s_{\alpha'}$ where $\alpha \neq \alpha'$:
\begin{equation} \label{c9}
\left\{ \begin{array}{rcl}
\displaystyle \left(z \psi_2 + {z (z-1) \over 2}\psi^2_1  \right) \eta_2 & = & \psi_2 \\*[4mm]
\displaystyle \left( z \psi_2 + {z (z-1) \over 2} \psi^2_1 \right) \frac{2(1-z\eta_2)}{z(z-1)} 
& = & \psi^2_1 \end{array} \right .
\end{equation}
Due to the constraint $\sum_\alpha q_{i,\alpha}=1$, one can obtain an identity by multiplying the 
first equation by $z$, and adding it to the second equation multiplied by $z(z-1)/2$. 
Hence, the two equations are not independent and $\psi_2$ can be solved. The value of $\psi_2$ 
depends only on $\eta_2$, the second moment of the $q$--distribution \cite{foot1}.

Working out the combinatorics of the higher orders, one finds the following general 
mathematical structure:
\begin{itemize}
\item At the $n$th order, there are as many equations as there are different partitions 
$\{n_1,n_2,\cdots,n_z \}$ that make $\sum_\alpha n_\alpha=n$. Permutations should not be considered as 
different because $\eta(\vec{q})$ is symmetric in its arguments.
\item One of these equations is dependent, as one can obtain an identity by adding the equations, 
after multiplication by appropriate factors.
\end{itemize}
For $z=2$, there are two third order equations, corresponding to the partitions 
$\{3,0\}$ and $\{2,1\}$, of which 
only one is independent. This means that $\psi_3$ can be solved as a function of $\eta_2$ (in 
appendix \ref{qmoments} we show that $\eta_3$ depends on $\eta_2$, for $z=2$). 
We run into problems at fourth order, where we have $\{4,0\}$, $\{3,1\}$ and $\{2,2\}$, and hence two 
a priori independent equations for one coefficient $\psi_4$. It turns out that the remaining 
equations are only identical if there is a relation between $\eta_4$ and $\eta_2$, namely
\begin{equation} \label{c10}
\eta_4=\frac{30\eta_2^2-11\eta_2+1}{16\eta_2-2}.
\end{equation}
In appendix \ref{qmoments}, it is shown that this relation is precisely obeyed by the class of $q$--
distributions Eq.(\ref{c2}) for which $\tilde{\psi}(s)$ was already solved.

\begin{figure}[ht]
\includegraphics[width=6.0cm]{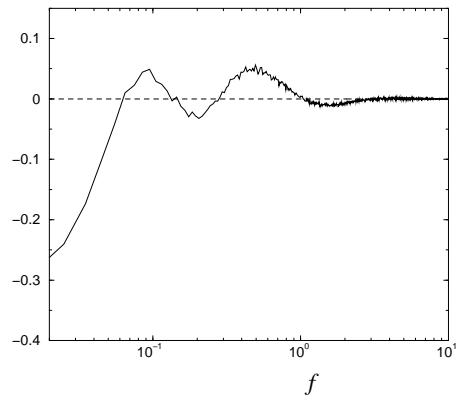}
\caption{Numerical simulation of a $q$--distribution with $q=0.1$ or $q=0.9$. The small 
deviation $p^{\rm mf}(f)-p^*(f)$ changes sign $4$ times.}
\label{wiggles}
\end{figure}

The fact that the expansion of $\tilde{\psi}(s)=[\tilde{p}^*(s)]^{1/z}$ only fails at fourth order 
implies that a mean field approximation, in which one explicitly assumes a product state, does give 
the exact results up to the third moment of $p^*(f)$. This is precisely the reason why 
the mean field solution $p^{\rm mf}(f)$ differs only marginally from the real solution. To be more 
precise, the deviation $p^{\rm mf}(f)-p^*(f)$ should change sign $4$ times, since it does not affect 
all moments lower than $\langle f^4 \rangle$. A careful inspection of the numerical results 
in \cite{cop} for a $q$--distribution in which $q=0.1$ or $q=0.9$ shows that these small 
``wiggles'' are indeed present. To magnify this effect, we show our simulation data in 
Fig. \ref{wiggles}.

For $z=3$, the problems already appear at third order. Since we have $\{3,0,0\}$, $\{2,1,0\}$ and 
$\{1,1,1\}$, we encounter two independent equations for $\psi_3$. 
Again, it turns out that the equations can be solved if there is an additional relation between 
the $q$--moments:
\begin{equation}\label{c10b}
\eta_3=\frac{15\eta_2^2-\eta_2}{9\eta_2+1}.
\end{equation}  

For $z>3$, there are two independent third order equations as well, originating from $\{3,0,0,0,\cdots\}$, 
$\{2,1,0,0,\cdots\}$ and $\{1,1,1,0,\cdots\}$. This problem can always be overcome by assuming a 
particular relation between the moments $\eta_3$ and $\eta_2$, corresponding to the special 
$q$--distributions of Eq.(\ref{c2}). Since at higher orders the number of equations per coefficient 
$\psi_n$ becomes increasingly high, there will be no other $q$--distributions than those of Eq.(\ref{c2}) 
that obey Eq.(\ref{b8}), and thus have an uncorrelated force distribution.

\section{Evolution of moments}\label{moments}
Now that we know that, in general, correlations do exist in the stationary force distributions, it 
is interesting to study the nature of these correlations. In this section, we 
write the evolution of the moments as master equations, along the lines of Ref. \cite{jac}. With this 
formalism, we will, in the next section, analyze the correlations 
by finding the stationary states of these master equations.

First, let us define the second moments of a distribution as
\begin{equation} \label{d1}
M_2(k)=\langle \, f_i f_{i+k} \, \rangle = \int d\vec{f}\,  f_i f_{i+k} \,P (\vec{f}). 
\end{equation}
We have reintroduced the site--index $i$, and $k$ is a displacement--vector in a layer. 
As the system is translationally invariant, these second moments depend only on the displacement $k$.
The recursion for these moments is obtained by combining Eq.(\ref{a2}) and Eq.(\ref{b1}) as
\begin{eqnarray} \label{d2}
M_2'(k)& =& \sum_{\alpha,\alpha'} \left( \int H(\vec{q}) \,d \vec{q} \,
q_{j,\alpha} \, q_{j+k+\alpha-\alpha',\alpha'} \right) \nonumber \\
&&\times \, \, M_2(k+\alpha-\alpha').
\end{eqnarray}
Using the overline notation for the $q$--averages again, Eq.(\ref{d2}) becomes
\begin{equation} \label{d3}
M_2'(k) = \sum_{\alpha,\alpha'} \overline{q_{j,\alpha} \,\, 
q_{j+k+\alpha-\alpha',\alpha'}} \,M_2(k+\alpha-\alpha').
\end{equation}
This relation is reveals from which points (in correlation--space) the moment 
$M_2'(k)$ receives a contribution during a recursion step. However, it is in fact easier to consider the 
opposite relation, revealing how much a moment contributes to correlation space points during recursion. 
The ``inverse'' of Eq.(\ref{d3}) becomes
\begin{equation} \label{d4}
M_2(k) \rightarrow \,\, \overline{q_{i,\alpha} \,q_{i+k, \alpha'}}
\, M_2'(k+\alpha'-\alpha) \quad\quad {\rm for}\,\,{\rm all}\quad \alpha, \alpha'.
\end{equation}
This latter relation allows for a master equation type formulation, as we may write it in the form
\begin{eqnarray} \label{d5}
M_2'(k) -  M_2 (k) = \nonumber \\
\sum_{\gamma} W_{\gamma} (k-\gamma)\, M_2 (k-\gamma) -W_{-\gamma} (k)\, M_2 (k).
\end{eqnarray} 
The transition rates are defined as
\begin{equation} \label{d6}
W_{\gamma} (k) = \overline{q_{i,\alpha} q_{i+k,\alpha'}},
\end{equation} 
with $\gamma$ determined by the set $\alpha, \alpha'$ as
\begin{equation} \label{d7}
\gamma = \alpha'-\alpha .
\end{equation} 
In the current problem, where we consider second order moments, the transition rates are 
particularly simple. If $k \neq 0$, the $q$--averages are independent, and will always give 
the value $1/z^2$ (this only holds for $q$--distributions of the type Eq.(\ref{a3})). 
If $k=0$, one encounters second moments of $\eta(\vec{q})$, as in Eq.(\ref{c6}). This leads 
to the following transition rates:
\begin{eqnarray} \label{d8}
k=0 &\Rightarrow& W_0(0)=\eta_2 \quad\quad W_{\gamma\neq0}(0)=\frac{1-z\eta_2}{z(z-1)}, \nonumber \\
k\neq 0&\Rightarrow& W_\gamma(k)=\frac{1}{z^2}.
\end{eqnarray}
So, the moments evolve in an anomalous diffusion process, with differing transition--rates at the 
origin. For a detailed discussion of the corresponding dynamics, see Ref. \cite{jac}. 
Note that this diffusion takes place in a $d-1$ dimensional 
space, as $\alpha$, and therefore also $\gamma$, is a displacement in a layer. In the remainder 
of this paper we use the bold notation $\mbox{\boldmath $\gamma$}$ whenever the displacement is really a vector. 

The advantage of this somewhat formal representation is that we can take it over 
to higher order moments without further ado. The generalization of the master equation for the $n$th 
order moments $M_n({\bf r})$ becomes:
\begin{eqnarray} \label{d10}
M_n' ({\bf r}) - M_n ({\bf r}) = \nonumber \\
\sum_{\gamma} W_{\gamma} 
({\bf r-} \mbox{\boldmath $\gamma$}) \, M_n ({\bf r}-\mbox{\boldmath $\gamma$}) - W_{-\gamma} ({\bf r})\, M_n ({\bf r}),
\end{eqnarray}
with the position indices ${\bf r}=(k_1,k_2,\cdots,k_{n-1})$, and the displacements 
$\mbox{\boldmath $\gamma$}$ defined as
\begin{equation} \label{d11}
\mbox{\boldmath $\gamma$} =(\alpha_1-\alpha,\alpha_2-\alpha, \cdots , \alpha_{n-1}-\alpha).
\end{equation}
The dimensionality of the diffusion process has now become $(n-1)(d-1)$. The transition rates can 
be calculated as
\begin{equation} \label{d12}
W_{\gamma} ({\bf r}) = \overline{q_{i,\alpha} q_{i+k_1,\alpha_1} \cdots 
q_{i+k_{n-1},\alpha_{n-1}}}.
\end{equation}
Analogous to the second 
moments, these transition rates are all $1/z^n$, as long as the indices of the position vector 
${\bf r}$ are not equal to zero nor coincide. However, the differing rates make the problem complicated, 
because one has to deal with different transition rates at special points, lines, planes etc. 
in the space of diffusion.

One can now study the correlations at infinite depth by finding stationary states of the master equation 
for the moments. As a first attempt to construct a stationary solution, i.e. 
$M_n'({\bf r})-M_n({\bf r}) =0$, one can try a detailed balance solution. 
Detailed balance means that there is no flow of ``probability'' from one point to another. In that 
case, all terms of the sum on the right hand side of Eq.(\ref{d10}) vanish individually, i.e. 
\begin{equation} \label{e1}
W_{-\gamma} ({\bf r})\, M_n ({\bf r}) =W_{\gamma} ({\bf r}-\mbox{\boldmath $\gamma$}) 
M_n ({\bf r}-\mbox{\boldmath $\gamma$}) \quad{\rm all} \quad {\bf r},\mbox{\boldmath $\gamma$}.
\end{equation}
This condition can also be formulated in terms of {\em elementary loops}, which are the smallest 
possible pathways from a point to itself. For all lattices in this study, these elementary loops are 
triangles, and we denote the three jump rates as $(a,b,c)$ or $(a',b',c')$ depending on the 
direction in which the loop is traversed. It is easily verified that the property
\begin{equation}\label{enew1}
abc=a'b'c'
\end{equation}
must be obeyed in {\em all} elementary loops in order to have a detailed balance solution. In the next section 
we show that correlations appear whenever the detailed balance conditions are not obeyed.

\section{Higher order correlations}\label{long--range}
In this section, we study the nature of the correlations for $q$--distributions of the 
type Eq.(\ref{a3}) that do not fall into the special class of Eq.(\ref{c2}). We first solve the stationary 
master equation for the second order moments, for which we already know that there are no 
correlations (section \ref{factorization}). For the triangular packing ($z=2$), correlations 
only show up at fourth order, and these fall off as $1/r^5$. For $z\geq3$, there are 
third order correlations that also decay with a power law; for the fcc packing ($z=3$) the decay is 
$1/r^4$. Finally, we provide a simple relation to calculate the various exponents.

\subsection{Second order moments: no correlations}
In order to get familiar with the structure of the master equations, we first 
consider the second order moments desribed by Eq.(\ref{d5}). Away from the origin $k=0$, all 
transition rates of Eq.(\ref{d8}) are identical. Therefore, the detailed balance condition Eq.(\ref{e1}) 
requires all $M_2(k\neq 0)$ to be identical. The value at the origin $M_2(0)$ has to obey a detailed balance 
condition for each $\gamma\neq0$, but these equations are identical for all $\gamma$ because the 
corresponding rates are the same. Putting $M_2(k\neq 0)=1$, one obtains the following stationary solution:
\begin{equation} \label{e2}
\langle \, f_i f_{i+k} \, \rangle= 
\left\{ \begin{array}{ll}
\displaystyle \frac{z-1}{z(1-z\eta_2)} & \mbox{, $k=0$} \\
\displaystyle 1 & \mbox{, $k\neq 0$}\end{array} \right. .
\end{equation}
This solution precisely describes an asymptotic state without any 2--point correlations, as the average 
of the product $\langle f_i f_j \rangle$ equals the product of the averages for all $i\neq j$. 
Of course, any multiple of Eq.(\ref{e2}), also forms a stationary solution of Eq.(\ref{d5}). However, 
these solutions are physically irrelevant in the thermodynamic limit, where the lattice size 
$\rightarrow \infty$ \cite{jac}. Moreover, we find that the asymptotic second force moment is 
solely determined by $z$ and $\eta_2$. For critical $q$--distributions one has $\eta_2=1/z$, leading 
to a diverging second moment.

\subsection{Third order moments}
The diffusion of third order moments $\langle f_if_{i+k}f_{i+l} \rangle$ takes place on a 
$2(d-1)$--dimensional lattice, since there are two free parameters $k$ and $l$ of dimension $d-1$. 
On this lattice, there are three {\em special subspaces}, namely $k=0$, $l=0$, and $k=l$, for which the 
transition rates of Eq.(\ref{d12}) differ from the {\em bulk}--value $1/z^3$. Moreover, the rates 
at the origin $k=l=0$ differ from both the bulk--rates and the rates on the special subspaces. 

\begin{figure}[ht]
\includegraphics[width=8.0cm]{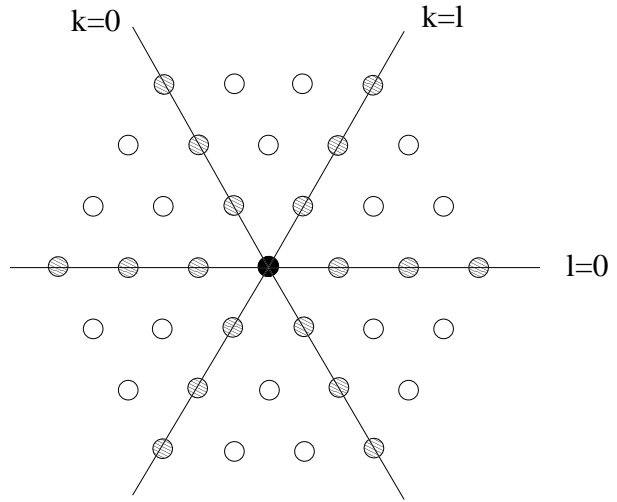}
\caption{Triangular packing: third order moments diffuse on a triangular lattice.}
\label{triang}
\end{figure}

Let us first consider the triangular packing ($z=2$), for which the third order moments diffuse on a 
$2$--dimensional lattice, with differing rates on three special lines. As these lines are all equivalent, 
it is natural to draw them at an angle of $120^{\circ}$, see Fig. \ref{triang}. 
We then obtain a triangular lattice, with transitions to six nearest neighbors and two {\em self jumps}, 
which are ``transitions'' to the same lattice site ($\mbox{\boldmath $\gamma$}=\bf{0}$). 
The detailed balance condition between a special line and the bulk is naturally identical to the 
second order condition, implying the same ratio as in Eq.(\ref{e2}). As the transition rates at 
the origin are again identical for each $\mbox{\boldmath $\gamma$}\neq {\bf 0}$ (because of symmetry), one can construct 
the following detailed balance solution:
\begin{equation} \label{e3}
\langle \, f_i f_{i+k} f_{i+l} \, \rangle= 
\left\{ \begin{array}{ll}
\displaystyle \frac{\eta_2}{(1-2\eta_2)^2} & \mbox{, origin} \\
\displaystyle \frac{1}{2(1-2\eta_2)} & \mbox{, lines} \\
1 & \mbox{, bulk}\end{array} \right. .
\end{equation}
This means that there are also no $3$--point correlations for $z=2$: at the origin we 
encounter $\langle f^3 \rangle$, on the lines we have 
$\langle f_i^2f_{i+k}\rangle=\langle f^2\rangle\langle f \rangle$, and in the 
bulk $\langle f_if_{i+k}f_{i+l}\rangle=\langle f\rangle^3$. It is easily checked 
that condition Eq.(\ref{enew1}) is indeed satisfied in every elementary loop. 

For the fcc packing ($z=3$), the third order moments diffuse on a $4$--dimensional lattice. Unlike the 
$z=2$ packing, it is not possible to construct a detailed balance solution in this case. First, we write the 
displacement vectors as $\mbox{\boldmath $\gamma$}=(\alpha'-\alpha,\alpha''-\alpha)=(\gamma_1,\gamma_2)$, where the $\alpha$'s 
and $\gamma$'s are 2--dimensional vectors (Fig. \ref{qmd}). One can jump away from the origin 
with two different rates, namely $\overline{q_1^2q_2}$ and $\overline{q_1q_2q_3}$. These rates correspond 
to $\gamma_1=\gamma_2$ (towards a special plane) and $\gamma_1\neq\gamma_2$ (into the bulk) respectively. 
Checking the detailed balance condition in the elementary triangle {\em origin--plane--bulk--origin}, 
it turns out that Eq.(\ref{enew1}) is only obeyed if $\eta_3$ and $\eta_2$ are related as in Eq.(\ref{c10b}). 
Of course, this is precisely the case for the class of Eq.(\ref{c2}) for which we know that asymptotic 
factorization occurs. In general, however, it is not possible to construct a detailed balance 
solution for the third order moments. In the next paragraph, we show that the absence of detailed 
balance indicates that there are force--correlations, that decay with a power law; in this case the 
decay is $1/r^4$.

\subsection{Fourth order moments}
\begin{figure}[ht]
\includegraphics[width=8.0cm]{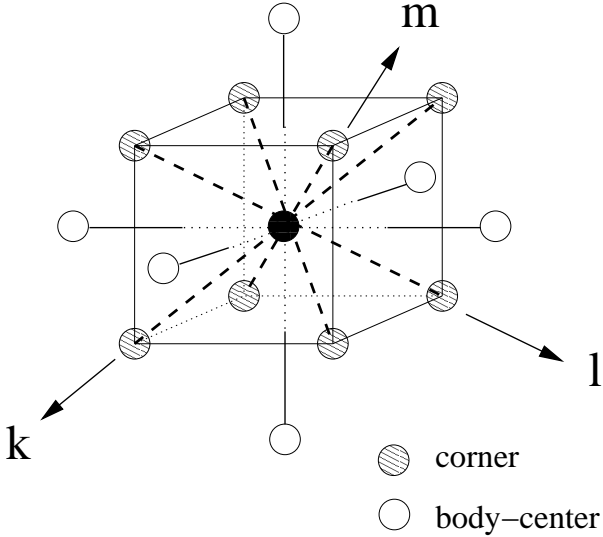}
\caption{Triangular packing: fourth order moments diffuse on a bcc lattice.}
\label{bcc}
\end{figure}

The fourth order moments $\langle f_if_{i+k}f_{i+l}f_{i+m} \rangle$ of the triangular packing diffuse on 
the bcc lattice depicted in Fig. \ref{bcc}. The three directions $k,l,m$ precisely define a 
bcc primitive cell \cite{ash}. There are now differing rates at the origin as well as on lines and planes 
for which one or more indices coincide or are equal to zero. The precise geometrical structure is 
explained in appendix \ref{appbcc}. There are now two a priori different 
directions away from the origin, that is to {\em corners} $\langle f^3_i f_{i+1} \rangle$ and to {\em body centers} 
$\langle f^2_i f^2_{i+1} \rangle$. Checking the loop condition Eq.(\ref{enew1}) for the loop 
{\em origin--corner--body center--origin}, one finds that it is only satisfied when $\eta_4$ and $\eta_2$ are 
related as in Eq.(\ref{c10}). 

The question that emerges is: What are the stationary solutions of the master equation, when the 
detailed balance condition is frustrated at the origin? 
To answer this question we first consider a simplified version of the bcc--problem, as a first order 
approximation. In this simple version, we assume that all jump rates are $1/z^4=1/16$, except at the origin 
where we distinguish between the two different directions. Although we neglect the differing rates 
on the special lines and planes, the loop condition is still frustrated in the elementary loop 
{\em origin--corner--body center--origin}. Using $\sum_{\gamma} W_{\gamma}({\bf r})=1$, we write 
the stationary master equation as
\begin{equation}\label{enew2}
M({\bf r})=\sum_{\gamma} W_{\gamma}({\bf r}- \mbox{\boldmath $\gamma$})M({\bf r}-\mbox{\boldmath $\gamma$}),
\end{equation}
or
\begin{equation}\label{enew3}
\left[ 1-2W_0({\bf r})\right]M({\bf r})  =
\sum_{\gamma \ne 0} W_{\gamma}({\bf r}- \mbox{\boldmath $\gamma$})M({\bf r}-\mbox{\boldmath $\gamma$}).
\end{equation}
This allows us to eliminate the two {\em self rates} $W_0$ by means of a simple transformation:
\begin{eqnarray}\label{enew4}
\hat{M}({\bf r})&=&\left[1-2W_0({\bf r}) \right] M({\bf r}),  \nonumber \\
\hat{W}_{\gamma}({\bf r})&=&W_{\gamma}({\bf r})/\left[1-2W_0({\bf r}) \right].
\end{eqnarray}
The sum over the new rates again adds up to unity and Eq.(\ref{enew3}) becomes 
\begin{equation}\label{enew5}
\hat{M}({\bf r})=\sum_{\gamma\neq 0} \hat{W}_{\gamma}({\bf r}-\mbox{\boldmath $\gamma$})
\hat{M}({\bf r}-\mbox{\boldmath $\gamma$}).
\end{equation}
Hence we can omit the self jumps by first solving the equation for the ``hatted'' variables, and then 
transforming back to $M({\bf r})$. As $M({\bf r})\rightarrow 1$ for large $r$, 
it is convenient to write
\begin{equation}\label{e4}
\hat{M}({\bf r})={7\over 8} \left[1+ \delta \hat{M}({\bf r}) \right].
\end{equation}
The quantity $\delta \hat{M}({\bf r})$ is in fact the appropriate measure for correlations \cite{foot2}. 
After eliminating the two self rates, all jump rates have become $1/14$, except at the origin where 
the rates to the $8$ corners $(c)$ can differ from the rates to the $6$ body centers $(b)$. 
We therefore have
\begin{equation}\label{enew6}
\hat{W}_{\gamma}({\bf r})=1/14 +\delta({\bf r}) \varepsilon_\gamma.
\end{equation}
The rates to the corners are denoted by $\varepsilon_c$ and those to the body centers by
$\varepsilon_b$. They fulfill the condition $8 \varepsilon_c+6 \varepsilon_b=0$.
This results in the following equation:
\begin{equation}\label{enew7}
\delta \hat{M}({\bf r})- {1\over 14} \sum_{\gamma \ne 0} \delta \hat{M}({\bf r}-\mbox{\boldmath $\gamma$})=
{8 \over 7}\hat{M}({\bf 0})\sum_{\gamma \ne 0} \varepsilon_\gamma \delta \left({\bf r}-\mbox{\boldmath $\gamma$} \right).
\end{equation}
Note that this is a discrete version of Poisson's equation: the LHS is a discrete Laplacian and 
the RHS, originating from deviating rates, acts as a multipole around the origin. 
This equation is solved in appendix \ref{appbcc} by a Fourier transformation, leading to
\begin{equation}\label{enew8}
\hat{M}({\bf r})={7 \over 8}+\hat{M}({\bf 0})\sum_{\bf k}\frac{E({\bf k})}{1-D({\bf k})}
\exp(-i{\bf k \cdot r}).
\end{equation}
The functions $D({\bf k})$ and $E({\bf k})$ are defined in appendix \ref{appbcc}; $1-D({\bf k})$ 
comes from the discrete Laplacian (in the continuum equation it would simply be $k^2$), 
$E({\bf k})$ is the Fourier transform of the source, and the sum over ${\bf k}$ is the inverse 
Fourier transformation running over the Brillouin Zone. 
The amplitude of the source $\hat{M}({\bf 0})$ can be obtained self--consistently, by setting 
${\bf r=0}$. This involves a complicated integral over the Brillouin Zone (BZ) 
of the bcc--lattice; the outcome, however, will be of the order unity. The large $r$ behavior 
of the correlations is determined by the small ${\bf k}$ behavior, so $E({\bf k})/(1-D(\bf{k}))$ 
has to be expanded around ${\bf k=0}$. The first term that gives a contribution is
\begin{eqnarray}\label{enew9}
{49 \varepsilon_c \over 24}\int \frac{d{\bf k}}{V_{\rm BZ}}\frac{\left
(k_x^2k_y^2+k_y^2k_z^2+k_z^2k_x^2\right) \exp(-i{\bf k\cdot r})}{k^2} \nonumber \\\
\simeq {343\varepsilon_c \over 32 \pi}\left[5\,\frac{x^2y^2+y^2z^2+z^2x^2}
{r^9}-{1\over r^5}\right].
\end{eqnarray}
The solution of Eq.(\ref{enew7}) decays as $1/r^5$; the terms $x^2y^2$ 
etc. give the proper angular dependence. 
This result can be directly understood from the analogy with electrostatics. The solution of Poisson's 
equation Eq.(\ref{enew7}) can be expanded in asymptotically vanishing spherical harmonics: 
$Y_{lm}/r^{l+1}$. The symmetry of the bcc lattice allows only harmonics with $l\ge 4$, leading to 
the observed $1/r^5$ decay.

So we find that the stationary master equation for the moments 
becomes a discrete Poisson's equation, and the presence of differing transition rates leads to a 
multipole source around the location of these rates, see Eq.(\ref{enew7}). However, this source is 
only ``active'' if there is no detailed balance, since detailed balance leads to trivial solutions 
like Eq.(\ref{e3}) \cite{foot4}. Keeping this in mind, 
let us now investigate the real fourth order problem, including the differing 
rates at the special lines and planes. We argue that the asymptotic value is still approached as 
$1/r^5$, but the amplitude of this field will be modified. Since there is no detailed balance, 
the differing rates at the lines and planes will act as sources as well. Their amplitudes, however, 
will decay with increasing distance, since the ``flow'' associated with the 
absence of detailed balance becomes zero at $r\rightarrow \infty$. The effect of the induced sources 
at the special lines and planes can be taken into account perturbatively. The first step is to only 
consider the effect of the origin, as we have done above. The second step would be to compute the 
strength of the sources at the lines and planes on the basis of the first order solution, and then to 
determine their function $E({\bf k})$ and recalculate the solution Eq.(\ref{enew8}). 
The induced sources around the origin basically lead to a modification of the strength 
$\hat{M}({\bf 0})$, but not of the asymptotic decay. However, the far away points at the lines and 
planes could modify the asymptotic decay. A closer inspection of the field of these sources shows 
that it is of order $1/r^7$, since the differing rates lead to a local Laplacian acting on the first 
order field decaying as $1/r^5$. Hence, every step of this perturbative calculation yields a leading 
term $1/r^5$; the amplitude changes in every step and its determination is a difficult problem indeed.

\subsection{Correlations for general $z$}
From the previous section, it is clear that correlations occur whenever the detailed balance condition 
is frustrated around the origin. The stationary master equation then becomes a discrete Poisson's equation 
in $(n-1)(d-1)$ dimensions, leading to correlations that decay with an integer power of the distance $r$. 
Following the derivation in appendix \ref{appbcc}, it is clear that the asymptotic behavior comes from 
the lowest non--isotropic term in $E({\bf k})$, since division by $1-D({\bf k})\approx k^2$ gives 
a singularity. The value of the exponent can be calculated as
\begin{equation}\label{enew10}
(n-1)(d-1)+order-2,
\end{equation}
where $(n-1)(d-1)$ is the dimensionality of the correlation space and $order$ is the order of the lowest 
non--isotropic terms in the expansion of $E({\bf k})$. Although this result is remarkably simple, the 
actual calculation of $E({\bf k})$ is not trivial, as it reflects the symmetries of the jump directions 
on the $(n-1)(d-1)$ dimensional lattice. Working out the $4$--dimensional lattice of the third order 
moments in the fcc packing, we find that $order=2$ and correlations vanish as $1/r^4$.

\section{Correlated $q$--distributions}\label{correlq}
So far, we have only discussed $q$--distributions of the type Eq.(\ref{a3}), for which there are 
no correlations between $q$--values at different sites. We have shown that, for these $q$--distributions, 
there are no asymptotic $2$--point force correlations. In this section we will demonstrate that even 
the smallest correlation between $q$--values at different sites induces $2$--point force correlations. 
We first solve the problem for arbitrary correlations in the triangular packing. Then, we study the 
fcc packing assuming only a nearest--neighbor $q$--correlation; this already leads to force 
correlations that decay as $1/r^6$.

\subsection{Triangular packing with arbitrary $q$--correlations}
In general, the (second order) transition rates are defined by Eq.(\ref{d6}). For $z=2$, 
the displacement vector $\alpha$ can only take two values, for which we conveniently choose 
$\pm \frac{1}{2}$. This allows us to write the transition rates as
\begin{eqnarray}\label{f1}
W_0(k)\,&=&\overline{q_{i,+\frac{1}{2}}\, q_{i+k,+\frac{1}{2}}} =
\overline{q_{i,-\frac{1}{2}}\, q_{i+k,-\frac{1}{2}}}, \nonumber \\
W_{+1}(k)&=&\overline{q_{i,-\frac{1}{2}}\, q_{i+k,+\frac{1}{2}}} \nonumber \\
&=& \overline{q_{i,-\frac{1}{2}}\, (1-q_{i+k,-\frac{1}{2}})} =1/2-W_0(k), \nonumber \\
W_{-1}(k)&=&\overline{q_{i,+\frac{1}{2}} q_{i+k,-\frac{1}{2}}} \nonumber \\ 
&=& \overline{q_{i,+\frac{1}{2}}\, (1-q_{i+k,+\frac{1}{2}})} =1/2-W_0(k).
\end{eqnarray}
Asymptotically $W_0(k)$ has to approach the value 1/4, for $q$--distributions without long--ranged 
correlations. As the second moments diffuse on a line, one can easily construct a detailed balance 
solution:
\begin{equation}\label{f2}
\left[1/2-W_0(k-1)\right] M(k-1) = \left[1/2-W_0(k)\right] M(k),
\end{equation}
or
\begin{equation}\label{f3}
M(k)=\frac{1/2-W_0(0)}{1/2-W_0(k)} M(0).
\end{equation}
This is the general form of the $2$--point force correlations $M(k)$ in the triangular packing, 
as a function of $W_0(k)$ that describes the $q$--correlations. 
One can draw two interesting conclusions from this result. 
First of all, there can only be an uncorrelated solution if $W_0(k)$ is constant (i.e. $1/4$) 
for each $k\neq 0$. This means that even the smallest $q$--correlations lead to force correlations. 
Secondly, the long distance behavior of the $2$--point force correlations is identical to that of the 
$2$--point $q$--correlations, following from the simplicity of Eq.(\ref{f3}).

\subsection{fcc packing with nearest neighbor $q$--correlations}
\begin{figure}[ht]
\includegraphics[width=5.0cm]{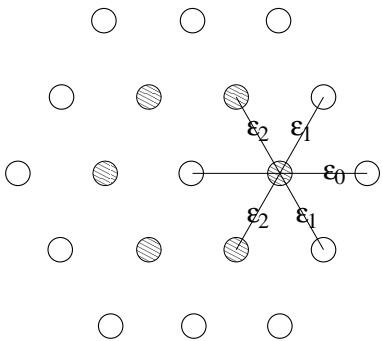}
\caption{fcc packing: second order moments diffuse on a triangular lattice. The ring around the 
origin has differing rates.}
\label{triangfcc}
\end{figure}

Unfortunately, the analysis is much more complicated for the fcc packing, whose second order 
moments live on the $2$--dimensional triangular lattice of Fig. \ref{triangfcc}. We therefore allow 
only correlations between $q$--values at neighboring sites. Remember that one can easily construct 
an uncorrelated solution $M({\bf r})$ for uncorrelated $q$--distributions, Eq.(\ref{e2}), since all detailed 
balance conditions at the origin are identical by symmetry. This still holds when there are 
nearest neighbor correlations. However, the detailed balance condition will now be frustrated on the 
ring of surrounding sites, as these are connected in four a priori different directions, see Fig. 
\ref{triangfcc}. 
In analogy to the problem discussed in the previous section, the stationary master equation for  
$\delta \hat{M}({\bf r})$ transforms into
\begin{equation}\label{f4}
\delta \hat{M}({\bf r}) - 1/6 \sum_{\gamma\neq {\bf 0}} \delta\hat{M}({\bf r}-\mbox{\boldmath $\gamma$}) = \rho({\bf r}).
\end{equation}
The ``charge density'' $\rho({\bf r})$ is only non-zero around the frustrated ring, see appendix 
\ref{appfcc}. 
Again, it is a discrete version of Poisson's equation, but now in $2$--dimensions. The solution can 
therefore be expanded in cylindrical harmonics, $\exp(in\phi)/r^n$, and the six--fold symmetry of the 
lattice requires $n\ge6$. The problem is again solved rigorously by Fourier transformation of 
Eq.(\ref{f4}). In appendix \ref{appfcc} we show that 
\begin{equation}\label{f5}
\delta\hat{M}({\bf r})\propto \frac{\cos(6\phi)}{r^6},
\end{equation}
which is in agreement with the simple electrostatic picture.

So, for the fcc packing, we find that even a nearest--neighbor $q$--correlation leads to 
$2$--point force correlations that decay with a power law. This algebraic decay is generic for $z\geq 3$ 
since any $q$--correlations lead to a master equation whose detailed balance relations cannot 
be solved around the origin.

\section{Discussion}\label{discussion}
We have studied force correlations in the $q$--model at infinite depth, 
for general $q$--distributions. The calculated correlation functions are rather unusual: 
for $q$--distributions of the type Eq.(\ref{a3}), correlations only show up at higher orders, 
and these correlations decay with a power of the distance. 
The only exceptions are the $q$--distributions given by Eq.(\ref{c2}), which do 
produce a factorized force distribution. The results for the triangular packing and the fcc packing are 
summarized in Table \ref{results}. As an example, consider two different sites $i$ and $i+k$ in a layer of the 
triangular packing. Since there are no correlations in the second and third order force moments, we find 
$\langle f_if_{i+k}\rangle=1$ and $\langle f_i^2f_{i+k}\rangle=\langle f^2\rangle$, independent of 
the distance $k$. However, the moments $\langle f_i^3f_{i+k}\rangle$ and $\langle f_i^2f_{i+k}^2\rangle$ 
are correlated and approach their asympotitic value as $1/k^5$. 
The fact that one has to go to higher orders to observe force correlations is the reason why numerical 
simulations only marginally differ from the mean field solutions \cite{cop}. The (single--site) mean field 
solutions $p^{\rm mf}(f)$ are correct up to the third order moments, for the triangular packing. 
This implies that $p^{\rm mf}(f)$ ``wiggles'' around the real solution $p^*(f)$; the deviation 
$p^{\rm mf}(f)-p^*(f)$ changes its sign $4$ times (Fig. \ref{wiggles}). 

Packings that have more than three $q$--values per site ($z\geq 3$) already have third order correlations. 
Also this time correlations only decay algebraically; for the fcc packing we find $1/r^4$. 
This algebraic decay can be understood from an analogy with electrostatics. 
The force moments evolve according to a master equation, and the corresponding stationary state is 
described by a discrete version of Poisson's equation. The ``source'' turns out to be a multipole 
around the origin, which is only active whenever the master equation has no simple detailed balance 
solution. The moments therefore approach their asymptotic (uncorrelated) values algebraically. The 
value of the exponent depends on the dimension of the correlation space $(n-1)(d-1)$, and on the symmetry of 
the multipole, see Eq.(\ref{enew10}).

Although in general correlations do exist, there is a special class of $q$--distributions, 
given by Eq.(\ref{c2}), for which there are no force correlations at all. 
This has been demonstrated by means of condition (\ref{b8}), which has a nice 
physical interpretation. It can be shown that the function $\psi(s)$ is the Laplace transform of the 
distribution of {\em interparticle forces} that live on the bonds connecting the particles: 
$v_{i,\alpha}=q_{i,\alpha}f_i$.
Although the $q$'s leaving a site are correlated (they have to add up to $1$), 
the corresponding $v_{i,\alpha}$ can become statistically independent. It is 
only when this miracle happens that the force distribution becomes a product state. Nevertheless, 
the $q$--distributions for which this is the case range from infinitely sharp 
($r\rightarrow \infty$) to almost critical ($r\rightarrow 0$). 

Finally, we found that there will be 2--point force correlations whenever the $q$--values of different 
sites are correlated. Even with only nearest neighbor $q$--correlations, the fcc packing has force 
correlations that vanish as $1/r^6$. Again, the triangular packing is less sensitive for correlations; 
the nature of the force correlations is identical to that of the $q$--correlations, Eq.(\ref{f3}).

{\bf Acknowledgements} The authors would like to thank Wim van Saarloos, Martin van Hecke and Martin 
Howard for stimulating discussions.

\begin{table}[ht]
\begin{center}
\begin{tabular}{|c||c|c|c|c|}
\hline 
& & & & \\*[-3mm]
packing & $n=2$ & $n=3$ &  $n=4$ & $n=2$, with $q$--corr. \\*[2mm]
\hline
& & & &\\*[-3mm]
triangular &line & triangular & bcc & line\\*[2mm]
($d=2$) &no corr. & no corr. &$1/r^5$ & like $q$--corr.  \\*[2mm]
\hline
& & & &\\*[-3mm]
fcc & triangular & $4$-dim. & $6$-dim.  & triangular \\*[2mm]
($d=3$) &no corr. & $1/r^4$ & -  & $1/r^6$ \\*[2mm]
\hline
\end{tabular} 
\caption{Summary of the results for the triangular packing ($z=2$; $d=2$) and the fcc packing 
($z=3$; $d=3$). The $n$th order force moments diffuse on a $(n-1)(d-1)$ dimensional lattice; the 
lattice structures are listed in the first row. The second row shows the nature of the 
corresponding force correlations in the stationary state.}
\label{results}
\end{center}
\end{table}

\appendix
\section{Moments of $q$--distributions}\label{qmoments}
This appendix is about the moments of the $q$--distri-butions, defined by 
\begin{equation} \label{A1}
\overline{q_1^{n_1}q_2^{n_2}...q_z^{n_z}} = \int \eta(\vec{q}) \delta 
\left( 1-\sum_\alpha q_\alpha \right) d \vec{q} \,\, q_1^{n_1}q_2^{n_2}...q_z^{n_z}.
\end{equation}
These different moments are not independent because of the $\delta$--constraint. As the distributions 
are normalised, the zeroth order moments are unity; the first order moments are, of course, all $1/z$. 
All second order moments, for which $\sum_i n_i=2$, can be described by only one free parameter. 
Defining $\eta_n$ as
\begin{equation} \label{A3}
\eta_n=\int \eta(\vec{q}) \delta \left( 1-\sum_\alpha q_\alpha \right) d\vec{q} \,\, q_1^n,
\end{equation}
one finds
\begin{eqnarray}\label{A4}
\sum_{i=1}^z \overline{q_1 q_i}=\eta_2 + (z-1)\overline{q_1 q_2} = \nonumber \\
\int \eta(\vec{q}) \delta \left( 1-\sum_\alpha q_\alpha \right) d\vec{q} \,\, q_1 \sum_{i=1}^z q_i= 1/z,
\end{eqnarray}
hence
\begin{equation}\label{A5}
\overline{q_1q_2}={1\over (z-1)}(1/z-\eta_2).
\end{equation}
From a similar argument, one can derive for the third order moments
\begin{equation}\label{A6}
\overline{q_1^3}=\eta_3 \quad\quad \overline{q_1^2 q_2}={1\over (z-1)}(\eta_2-\eta_3).
\end{equation}
For $z=2$ there is even a relation between $\eta_3$ and $\eta_2$:
\begin{equation}\label{A7}
1=\sum_{ijk}\overline{q_iq_jq_k}=2\eta_3+6\overline{q_1^2q_2} \quad\Longrightarrow \quad 
\eta_3={3\over 2}\eta_2 -1/4.
\end{equation}
For $z=3$, there is an additional third moment, namely 
\begin{eqnarray}\label{A8}
1=\sum_{ijk}\overline{q_iq_jq_k}=3\eta_3+18\overline{q_1^2q_2}+6\overline{q_1q_2q_3} \nonumber \\
\Longrightarrow \quad \overline{q_1q_2q_3}={1\over 6}(1-9\eta_2+6\eta_3).
\end{eqnarray}
The extension to higher orders and higher $z$ is straightforward.

For the special class of $\eta(\vec{q})$ defined in Eq.(\ref{c2}), one can calculate the moments 
$\eta_n$ from a generalization of Eq.(\ref{zinnjust}) \cite{zinn}
\begin{equation}\label{A9}
\eta_n=\frac{\Gamma(zr)\Gamma(r+n)}{\Gamma(r)\Gamma(zr+n)}.
\end{equation}
In order to show that Eq.(\ref{c10}) is indeed obeyed by the special class (with $z=2$), we first 
invert Eq.(\ref{A9}) for $n=2$
\begin{equation}\label{A10}
r=\frac{1-2\eta_2}{4\eta_2-1}.
\end{equation}
From this one can calculate $\eta_4$ as a function of $\eta_2$, which precisely results in 
Eq.(\ref{c10}). A similar inversion for $z=3$ leads to 
\begin{equation}\label{A11}
r=\frac{1-3\eta_2}{9\eta_2-1},
\end{equation}
from which one derives Eq.(\ref{c10b}).

\section{The bcc lattice}\label{appbcc}
In the triangular packing, the fourth order force moments $\langle f_if_{i+k}f_{i+l}f_{i+m} \rangle$ 
diffuse on the bcc lattice of Fig. \ref{bcc}, with differing jump rates on special lines and planes. 
In this appendix, we list these rates explicitly and we solve the corresponding stationary master 
equation. 

The jump rates can be calculated from 
\begin{equation} \label{B1}
W_{\gamma} (k,l,m) = \overline{q_{i,\alpha}\, q_{i+k,\alpha'}\,q_{i+l,\alpha''}\,q_{i+k,\alpha'''}},
\end{equation} 
with the $z^4=16$ jump directions
\begin{equation} \label{B2}
\mbox{\boldmath $\gamma$} =(\alpha'-\alpha,\alpha''-\alpha,\alpha'''-\alpha).
\end{equation}
As $\alpha$ can take the values $\pm {1\over 2}$, there are two {\em self rates} 
for which all $\alpha$'s are the same. As a consequence, there are $14$ outgoing 
directions, namely $\pm (1,0,0)$, $\pm (1,1,1)$, and $\pm (1,1,0)$ plus their permutations. The first 
two are directions for which three of the four $\alpha$'s are equal, and they correspond to the 
{\em corners} of Fig. \ref{bcc}; the third represents the jumps towards the {\em body centers}. 
If all position indices in Eq.(\ref{B1}) are different, the transition rates are simply $1/z^4=1/16$. 
On the special lines and planes where one or more position indices coincide, we encounter differing rates. 
The geometry of the problem is summerized in Table \ref{table}. 
\begin{table}[ht]
\begin{center}
\begin{tabular}{|cc|ccccc|}
\hline 
& & & & & &\\*[-3mm]
&from $\setminus$ to & 
$ (0,0,0) $ & $(k,0,0)$ &  $(k,k,0) $ & $(k,l,0)$ & $(k,l,m) $  \\*[2mm]
\hline
& & & & & &\\*[-3mm]
origin& $(0,0,0)$ & $\overline{q_1^4}$ & $\overline{q_1^3q_2}$ & $\overline{q_1^2q_2^2}$ & 
$-$ & $-$ \\*[2mm]
line (c) & \begin{tabular}{c} $(k,0,0)$ \\ $(k,k,k)$ \end{tabular}& 
${1\over 2}\overline{q_1^3}$ & ${1\over 2}\overline{q_1^3}$ 
& ${1\over 2}\overline{q_1^2q_2}$ & ${1\over 2}\overline{q_1^2q_2}$ & $-$ \\*[3mm]
line (b) & $(k,k,0)$ & $(\overline{q_1^2})^2$  & $\overline{q_1^2}\,\overline{q_1q_2}$ & 
$(\overline{q_1^2})^2$  &
$\overline{q_1^2}\,\overline{q_1q_2}$ &$(\overline{q_1q_2})^2$ \\*[1mm]
plane& \begin{tabular}{c} $(k,l,0)$\\ $(k,k,l)$ \end{tabular}& $-$ & 
${1\over 4}\overline{q_1^2}$ & ${1\over 4}\overline{q_1^2}$  & 
${1\over 4}\overline{q_1^2}$ & ${1\over 4}\overline{q_1q_2}$ \\*[2mm]
bulk& $(k,l,m)$ & $-$ & $-$ & ${1\over 16}$ &
${1\over 16}$ & ${1\over 16}$ \\*[2mm]
\hline
\end{tabular} 
\caption{The transition rates $W_{\gamma} ({\bf r})$ for the fourth order master equation.}
 \label{table}
\end{center}
\end{table}

From this table we deduce the rates $\varepsilon_c$ to the corners and $\varepsilon_b$ 
to the body centers, which occur in relation (\ref{enew6}). We find
\begin{equation} \label{B3}
\varepsilon_c = {\overline{q^3_1 q_2} \over 1 - 2 \overline{q^4_1}} - {1 \over 14}, \quad \quad \quad
\varepsilon_b = {\overline{q^2_1 q^2_2} \over 1 - 2 \overline{q^4_1}} - {1 \over 14},
\end{equation} 
and one easily verifies from the property $q_1 + q_2 =1$ that the relation 
$8 \varepsilon_c + 6 \varepsilon_b = 0$ holds. In general, the rates do not obey the detailed 
balance condition Eq.(\ref{enew1}) in the elementary loop {\em origin--corner--body center--origin}. 
Keeping only the rates in this loop as deviations from the bulk leads to equation (\ref{enew7}).
For the definition of the two functions $E ({\bf k})$ and $D {\bf k})$ we introduce two auxiliary functions:
one for the contribution of the corners
\begin{eqnarray} \label{B4}
\tilde{E}_c ({\bf k}) &=& {1\over4} \left[\cos {k_x +k_y + k_z \over 2} +
 \cos {k_x -k_y + k_z \over 2} \right. \nonumber \\ 
&& \left. + \cos {k_x +k_y - k_z \over 2}  + \cos {k_x -k_y - k_z \over 2} \right] \nonumber \\
&&
\end{eqnarray} 
and one related to the body centers
\begin{equation} \label{B5}
\tilde{E}_b ({\bf k}) = {1\over3} (\cos k_x + \cos k_y + \cos k_z).
\end{equation} 
The two functions $\tilde{D} ({\bf k})$ and $\tilde{E} ({\bf k})$ are then given as
\begin{eqnarray} \label{B6}
\tilde{D} ({\bf k}) &=& {4\over7} \tilde{E}_c ({\bf k}) + {3\over7} \tilde{E}_b ({\bf k}) \nonumber \\
\tilde{E} ({\bf k}) &=& \varepsilon [\, \tilde{E}_c ({\bf k}) -\tilde{E}_b ({\bf k}) \,],
\end{eqnarray}
with $\varepsilon = 8 \varepsilon_c = - 6 \varepsilon_b$. 

For the large ${\bf r}$ behavior we need the expansions for small $k$. One finds
\begin{equation} \label{B7}
\tilde{E}_c ({\bf k}) = 1 - {1\over 8} k^2 + {1\over384} [\,k^4 + 4 
(k^2_x k^2_y +k^2_y k^2_z +k^2_z k^2_x )\,]  + \cdots
\end{equation} 
and 
\begin{equation} \label{B8}
\tilde{E}_b ({\bf k}) = 1- {1\over6} k^2 + {1\over72} [\,k^4 - 
(k^2_x k^2_y +k^2_y k^2_z +k^2_z k^2_x )\,]  + \cdots
\end{equation} 
From these expressions one derives the expansion
\begin{eqnarray} \label{B9}
{\tilde{E}  ({\bf k}) \over 1 - \tilde{D} ({\bf k}) } &=&
{7\epsilon \over 24} \left( 1 - {7 \over 32}k^2 \right. \nonumber \\
&& \left. + {7 \over 8} {k^2_x k^2_y +k^2_y k^2_z +k^2_z k^2_x  \over k^2} + \cdots \right).
\end{eqnarray} 
The first two terms in the expansion are regular and thus give rise to short range contributions. 
The last term leads to the asymptotic behavior, by means of the inverse Fourier transform 
\begin{equation} \label{B10}
\int \frac{d{\bf k}}{V_{\rm BZ}}\frac{\left
(k_x^2k_y^2+k_y^2k_z^2+k_z^2k_x^2\right) \exp(-i{\bf k\cdot r})}{k^2}.
\end{equation}
This integral can be evaluated by differentiation of the well--known
\begin{equation} \label{B11}
\int \frac{d{\bf k}}{V_{\rm BZ}}\frac{\exp(-i{\bf k\cdot r})}{k^2}\simeq \frac{1}{4\pi r},
\end{equation}
where a factor $k_x$ in Eq.(\ref{B10}) corresponds to applying $\partial / \partial x$. 
This leads to expression (\ref{enew9}).

\section{$q$--correlations in the fcc packing}\label{appfcc}

In Eq. (\ref{f4}) we formulated the problem for the second moments in the fcc packing with 
nearest--neighbor $q$-correlations. The ``charge density'' $\rho ({\bf r})$ on the right hand side
of the equation is the product of the moment $\hat{M}(\mbox{\boldmath $\gamma$})$, referring to the neighbors
of the origin (all are the same by symmetry), with a function whose Fourier transform is given by
\begin{equation} \label{C1}
\tilde{E} ({\bf k}) = \sum_{\gamma',\gamma} w_{\gamma - \gamma'} 
\exp [i {\bf k} \cdot (\mbox{\boldmath $\gamma$} + \mbox{\boldmath $\gamma$}')].
\end{equation} 
The $w_{\gamma -\gamma'}$ are the deviations from the bulk transition rates $1/6$. These are only 
non--zero for the ring of nearest neighbors around the origin shown in Fig. \ref{triangfcc}:
\begin{eqnarray} \label{C2}
w_0 &=& - \varepsilon_0 , \quad\quad\quad\,\,\,  w_1 = w_5 = - \varepsilon_1, \nonumber \\ 
w_2 &=& w_4 = - \varepsilon_2, \quad  w_3 = \varepsilon_0 + 2\varepsilon_1 + 2\varepsilon_2.
\end{eqnarray} 
The equalities reflect the symmetry of the triangular lattice.
Inserting Eq.(\ref{C1}) into the Fourier transform of Eq.(\ref{f4}) leads to
\begin{equation} \label{C3}
\hat{M} ({\bf r}) = 2/3 +  \hat{M}(\mbox{\boldmath $\gamma$})\sum_{\bf k} { \tilde{E}({\bf k}) \over 1-\tilde{D} ({\bf k})}
\,\exp (- i {\bf k} \cdot {\bf r}).
\end{equation}
The consistency equation for $\hat{M}(\mbox{\boldmath $\gamma$})$ follows by taking ${\bf r}$ as one of the nearest 
neighbors of the origin. The function $\tilde{D} ({\bf k})$ is given by
\begin{equation} \label{C4}
\tilde{D} ({\bf k}) = {1 \over 3} \left( \cos k_x + 
\cos {k_x +\sqrt{3}k_y \over 2} + 
\cos {k_x -\sqrt{3}k_y \over 2} \right),
\end{equation}
and $\tilde{E} ({\bf k})$ can be expressed as
\begin{equation} \label{C5}
\tilde{E} ({\bf k})/6 = \varepsilon_0 [1 - \tilde{D} (2 {\bf k})] + 
2\varepsilon_1 [1 - \tilde{D}'({\bf k})] + 2\varepsilon_2 [1 - \tilde{D} ({\bf k})],
\end{equation}
with the new function 
\begin{equation} \label{C6}
\tilde{D}'({\bf k}) = \tilde{D}(\sqrt{3}\, k_y,\sqrt{3}\, k_x).
\end{equation}  
For the asymptotic behavior of $\hat{M} ({\bf r})$ we must make an expansion of 
$\tilde{E} ({\bf k})/(1-\tilde{D} ({\bf k}))$. For the first two terms we find
\begin{eqnarray} \label{C6a}
\frac{1 - \tilde{D} (2 {\bf k})}{1 -\tilde{D} ({\bf k})}&=&\, 
4\,\left( 1 - {3 \over 16}k^2 + {3\over 256} k^4 \right. \nonumber \\
&+&  \left. {1\over 192}\frac{k_x^6-6k_x^4k_y^2+9k_x^2k_y^4}{k^2}+\cdots \right),
\end{eqnarray}
\begin{eqnarray} \label{C6b}
\frac{1 - \tilde{D}' ({\bf k})}{1 -\tilde{D} ({\bf k})}&=& 
3\,\left( 1 - {1 \over 8}k^2 + {1\over 128} k^4 \right. \nonumber \\
&& + \left.{1\over 288}\frac{k_x^6-6k_x^4k_y^2+9k_x^2k_y^4}{k^2}+\cdots \right), \nonumber \\
\end{eqnarray}
and the third term is simply a constant. The asymptotic behavior is given by Fourier inversion 
of the first singular term in ${\bf k}$, i.e.
\begin{eqnarray} \label{C7}
\int {d{\bf k} \over V_{BZ}} \,\frac{\left(k_x^6-6k_x^4k_y^2+9k_x^2k_y^4 \right)
\, \exp (-i {\bf k \cdot r})}{k^2} \nonumber \\
\simeq {960 \over \pi}\frac{\cos(6\phi)}{r^6}.
\end{eqnarray}
This integral can be obtained by differentiation of 
\begin{equation} \label{C12}
\int {d {\bf k} \over V_{BZ}} {\exp (- i {\bf k} \cdot {\bf r}) \over k^2} 
\simeq {\log (L/r) \over 2 \pi },
\end{equation}
where $L$ is the size of the system.


\begin{thebibliography}{99}
\bibitem{foot3} The limit $r \rightarrow 0$ is a route towards the critical 
distribution because the $q$--values become increasingly concentrated around $0$ and $1$. 
Although the distribution of Eq.(\ref{c2}) is not normalisable for $r=0$, all higher 
moments approach the moments of the critical distribution for $r \rightarrow 0$, see appendix 
\ref{qmoments}. The corresponding force distributions approach $1/f$ with a cut--off for large 
forces.
\bibitem{foot1} The remaining equation for $\psi_2$ cannot be solved if $\eta_2=1/z$, i.e. 
for the critical $q$--distribution. As the corresponding force distribution has diverging moments, 
one cannot even Taylor expand $\tilde{p}^*(s)$ and $\tilde{\psi}(s)$.
\bibitem{foot4} A {\it discrete} Poisson's equation with a multipole source term can indeed have a 
short--ranged solution. This is the case if $E({\bf k})=K({\bf k})(1-D(\bf{k}))$, and $K({\bf k})$ 
is a regular function.
\end{thebibliography}
\end{document}